\documentclass[aps,reprint,prl, superscriptaddress, footinbib]{revtex4-1}

\usepackage{color}
\usepackage{amsmath} 
\usepackage{amssymb}
\usepackage{graphicx}% Include figure files
\usepackage[caption=false]{subfig} %multi figures
\usepackage{dcolumn}% Align table columns on decimal point
\usepackage{bm}% bold math
\usepackage{hyperref}% add hypertext capabilities
\usepackage{verbatim} 

\newcommand{\phicr}{\phi_\mathrm{cr}}

\newcommand{\Tscr}{T^*_\mathrm{cr}}

\newcommand{\GM}{\hat{G}_\mathrm{M}}

\begin{document}

\title{Sequence-specific polyampholyte phase separation in 
membraneless organelles}
\author{Yi-Hsuan Lin}
\affiliation{Department of Biochemistry, University of Toronto, 1 King's College Circle, Toronto, Ontario M5S 1A8, Canada}
\affiliation{Molecular Structure and Function Program, Hospital for Sick Children, 686 Bay Street, Toronto, ON M5G 0A4, Canada}
\author{Julie D. Forman-Kay}
\affiliation{Molecular Structure and Function Program, Hospital for Sick Children, 686 Bay Street, Toronto, ON M5G 0A4, Canada}
\affiliation{Department of Biochemistry, University of Toronto, 1 King's College Circle, Toronto, Ontario M5S 1A8, Canada}
\author{Hue Sun Chan}
\affiliation{Department of Biochemistry, University of Toronto, 1 King's College Circle, Toronto, Ontario M5S 1A8, Canada}
\affiliation{Department of Molecular Genetics, University of Toronto, Toronto, 1 King's College Circle, Ontario M5S 1A8, Canada}

\date{2016/09/17}	

\begin{abstract}

Liquid-liquid phase separation of charge/aromatic-enriched intrinsically 
disordered proteins (IDPs) is critical in the biological function of 
membraneless organelles. Much of the physics of this recent discovery remains 
to be elucidated. Here we present a theory in the random phase approximation 
to account for electrostatic effects in polyampholyte phase separations,
yielding predictions consistent with recent experiments on the IDP Ddx4. 
The theory is applicable to any charge pattern and thus provides a general 
analytical framework for studying sequence dependence of IDP phase separation. 

\end{abstract}


\maketitle

%\section{Introduction}

The biological function of proteins has long been associated with their 
ordered, often globular, structures. It is now clear, however, 
that many critical cellular functions---in signaling and cell-cycle regulation
in particular---are performed by intrinsically disordered proteins (IDPs) 
that lack a unique fold.  IDPs are depleted in hydrophobic but enriched 
in charged and polar amino 
acids~\cite{Uversky00,tompa12,FormanKay13,vanderLee14,Chen15}. 
Many IDPs are polyampholytes with both positively 
and negatively charged monomers~\cite{Higgs91, Dobrynin04}. Recently,
some polyampholytic IDPs were found to function not only as individual
molecules but also collectively by undergoing phase 
separation on a mesoscopic length scale to form 
condensed liquid-phase non-amyloidogenic IDP-rich droplets that may 
encompass RNA and other 
biomolecules~\cite{Brangwynne2009, Rosen12, McKnight12, Lee13, Nott15}. 
This phase behavior is the basis of functional membraneless 
organelles in the cell. Without a membrane, these organelles can 
respond rapidly to environmental stimuli. They are critical for cellular 
integrity, homeostasis~\cite{Nott15, Dundr10}, 
and the spatial-temporal control of gene regulation and cell 
growth~\cite{Lee13, wright14, julicher14, Nott15, ElbaumGarfinkle15, Brangwynne15}. 
In view of the biological/biomedical importance of this newly discovered
phenomenon, insights into its physics would be valuable. 

In line with charge effects
on the size of individual IDP molecules~\cite{Mao10, Marsh10, Muller10, Das13}, 
electrostatics figured prominently in 
computational~\cite{Ruff15} and experimental~\cite{Nott15, Quiroz15} analysis 
of IDP phase separation. Polymer theory emphasized the sensitivity 
of polyampholyte phase behavior to charge pattern~\cite{Higgs91, Dobrynin04},
but only a few simple patterns were 
considered~\cite{Gonzalez94, Cheong05, Jiang06}.
No systematic approach has been put forth to apply those ideas to 
understand phase behaviors of genetically coded proteins.
Inasmuch as IDP phase separation is concerned, the theoretical discussion
to date remains at the mean-field 
level~\cite{Lee13, julicher14, Nott15, Brangwynne15},
which precludes sequence dependence to be addressed. 
In this Letter, we take a step forward by developing an analytical theory 
for sequence-specific electrostatics in polyampholyte phase separation,
aiming to synergize theory and experiment and lay the groundwork for 
further theoretical advances.

While our theory is applicable to any charge sequence,
an impetus for our effort was the recent experiments
on the RNA helicase Ddx4, the N-terminal of which is an IDP, 
that can undergo {\it in vitro} and in cell phase 
separation under ambient conditions~\cite{Nott15}. Ddx4 
is essential for the assembly and maintenance of the related nuage in mammals, 
P-granules in worms, as well as pole plasm and polar granules in 
flies~\cite{Nott15, Liang94}. The wildtype sequence of the 
residue 1--236 N-terminal disordered region of Ddx4 (termed Ddx4$^{\rm N1}$) 
may be seen by sliding-window averaging as a series of alternating charge 
blocks abounding with negatively charged aspartic (D) and glutamic (E) acids 
and positively charged arginines (R) and lysines (K) \cite{Nott15}. 
In the absence of this block-charge pattern, 
a charge-scrambled mutant Ddx4$^{\rm N1}$CS 
does not phase separate~\cite{Nott15}. Thus the sequence-specific charge
pattern---not total positive and total negative charges per se---is 
crucial for Ddx4 phase behavior.

Since it is challenging to synthesize
nonbiological polyampholytes with specific charge patterns~\cite{Dobrynin04},
theories have focused on the quenched ensemble 
average of random sequences~\cite{Higgs91, Wittmer93} 
or limited to diblock~\cite{Gonzalez94} 
or at most four-block charge patterns~\cite{Cheong05, Jiang06}. 
In contrast, a high diversity of protein sequences are readily synthesized by
the cellular machinery. In view of this expanded experimental horizon, 
the availability of a large repertoire of IDP sequences 
for phase separation studies is foreseeable, with the new physics that
is likely to ensue.
With this in mind, we present below a random phase approximation (RPA) 
theory~\cite{Borue88, Wittmer93, Gonzalez94,Mahdi00, Ermoshkin03a, Ermoshkin04} 
for any charge pattern and, as an example, apply our theory to elucidate 
Ddx4 phase behavior.

As in prior applications of RPA, the particle density in our theory 
is assumed to be rather homogeneous to permit an approximate account
based solely on two-body correlations of density fluctuation~\cite{Borue88}. 
Electrostatic 
free energy is similarly approximated by a Gaussian integral over charge 
fluctuations. Other limitations of RPA, such as in its treatment of
short-range electrostatic interactions, are well documented
(e.g.~\cite{Borue88,Mahdi00}). 
Approximations notwithstanding, RPA represents
a significant improvement, especially amidst the current interest
in IDP phase separation \cite{julicher14, Brangwynne15}, over 
Debye-H\"{u}ckel theory for 
biological coacervation~\cite{Overbeek57} in that RPA 
embodies chain connectivity~\cite{Borue88,Mahdi00}
and hence allows for an explicit account of the charge 
pattern along the chain sequence~\cite{Wittmer93,Gonzalez94}.

%\section{The RPA theory}

Based on previous approaches~\cite{Wittmer93, Gonzalez94}, 
our theory is for a system of aqueous polyampholytes, small
counterions, and salt. In contrast to self-repelling, strongly charged 
polyelectrolytes that necessitate a modified 
RPA~\cite{Mahdi00, Ermoshkin03b}, 
we focus on polyampholytes that are nearly neutral, i.e., with
approximately equal numbers of acidic and basic residues 
(e.g. Ddx4$^\text{N1}$ and Ddx4$^\text{N1}$CS). Each 
polyampholyte consists of $N$ monomers 
(amino acid residues) with 
charges  $\{\sigma_i\} = \{ \sigma_1,\sigma_2,\dots,\sigma_N \}$, where 
$\sigma_i$ is in units of the electronic charge $e$. 
The counterions and salt are monovalent in this formulation.
The densities of monomers, counterions, and salt are denoted, respectively, 
by $\rho_m$, $\rho_c$, and $\rho_s$. The number of monovalent counterions 
is taken to be equal to the net charge of polyampholytes, viz.,
$\rho_c = \rho_m\left|\sum_i \sigma_i\right|/N$.
The free energy $F$ of our system per volume $V$ in units of $k_{\rm B}T$ 
is 
\begin{equation}
f \equiv \frac{F}{V k_{\rm B}T} = -s + f_{\rm el}  \; ,
\end{equation}
where $k_{\rm B}$ is Boltzmann's constant and $T$ is absolute temperature,
$s$ is the entropic contribution, and $f_{\rm el}$ accounts for electrostatic
interactions. As IDPs have few hydrophobic residues, as a first step in
our development, we assume that electrostatics is the dominant enthalpic 
contribution while neglecting presumably weaker short-range attractive 
forces. For simplicity, all monomers of the polyampholytes as well as the 
counterions and salt ions are taken to be of equal size with length scale $a$.
The entropic term that accounts for excluded volume follows directly 
from Flory-Huggins (FH) theory~\cite{Flory53, deGennes79, chandill94}:
\begin{equation}
\begin{aligned}
-s a^3 = & \frac{\phi_m}{N} \ln \phi_m + \phi_c \ln\left(\phi_c \right) +  
2\phi_s \ln(\phi_s)  \\
	&+ (1\!-\!\phi_m\!-\!\phi_c\!-\!2\phi_s)\ln(1\!-\!\phi_m\!-
\!\phi_c\!-\!2\phi_s),
	\label{eq:entropy}
\end{aligned}
\end{equation}
where $\phi_m$, $\phi_c$, $\phi_s$ are, respectively, the volume ratios 
$\rho_m a^3$, $\rho_c a^3$, and $\rho_s a^3$. 
A uniform, phase-independent $\phi_s$ is assumed here for simplicity
as previous work suggested that change in salt concentration is 
insignificant upon biological coacervation~\cite{Overbeek57}.
The $f_{\rm el}$ term for electrostatics is computed by 
RPA~\cite{Wittmer93, Mahdi00, Ermoshkin03a, Ermoshkin04}:
\begin{equation}
f_{\rm el} = \frac{1}{2}\int \frac{d^3 k}{(2\pi)^3} 
\left\{\ln[\det(1 + \hat{G}_k\hat{U}_k)] - \mathrm{Tr}
(\hat{\rho}\;\hat{U}_k) \right\},
	\label{eq:f_el}
\end{equation}
where $\hat{G}_k$, $\hat{U}_k$, and $\hat{\rho}$ are $(N+2)\times(N+2)$ 
matrices. The logarithmic term containing $\hat{G}_k$ for basic, ``bare'' 
density correlation without electrostatic effects and $\hat{U}_k$ for 
Coulombic interactions, both defined for the reciprocal $k$-space, is 
a result of standard Gaussian integrations in RPA theory.
The second trace term subtracts the self electrostatic energy 
for all charged densities to eliminate the unphysical divergence of the 
first term for $k\!\to\!\infty$~\cite{Wittmer93, Ermoshkin03a, Ermoshkin04}. 
The density matrix
\begin{equation} 
\hat{\rho} = \left(
\begin{array}{cc}
(\rho_m/N) \hat{I}_N & 0 \\
0 & \hat{\rho}_I
\end{array}\right)
\end{equation}
is a diagonal matrix, with $\hat{I}_N$ being the $N$-dimensional identity 
and $\hat{\rho}_I$ a $2\times 2$ diagonal matrix for the 
positive and negative monovalent ions, namely
$((\hat{\rho}_I)_{11}, (\hat{\rho}_I)_{22})=(\rho_s\!+\!\rho_c, \rho_s)$ or
$(\rho_s, \rho_s\!+\!\rho_c)$ when the net polyampholyte charge is, 
respectively, negative or positive.

The bare correlation matrix $\hat{G}_k$ combines the monomer-monomer 
correlation for a Gaussian chain and the density matrix for the
small monovalent ions:
\begin{equation} 
\hat{G}_k = \left(
\begin{array}{cc}
(\rho_m/N) \GM(k) & 0 \\
0 & \hat{\rho}_I
\end{array}\right),
\end{equation}
where $\GM(k)$ is the $N\times N$ matrix for Gaussian chains, with
elements $\GM(k)_{ij} = \exp(-(ka)^2|i-j|/6)$~\cite{DoiEdwardsbook}. 

The interaction matrix $\hat{U}_k$ is the Fourier transform of 
the Coulomb potential with a short-range physical cutoff on the scale of 
monomer size, $U(r) = l_B(1-e^{-r/a})/r$,
which in $k$-space becomes~\cite{Ermoshkin03a, Ermoshkin04}
\begin{equation}
\hat{U}_k = \frac{4\pi l_B}{k^2[1+(ka)^2]}| 
q\rangle \langle q | \equiv \lambda(k)| q\rangle \langle q |  .  
\end{equation}
Here $l_B = e^2/(4\pi\epsilon_0\epsilon k_{\rm B} T)$ is the Bjerrum length,
$\epsilon_0$ is vacuum permittivity,
$\epsilon$ is the dielectric constant, $| q \rangle$ is the column vector 
for the charges of the monomers and monovalent ions, and 
$\langle q | \equiv | q \rangle^{\rm T}$ is the transposed row vector,
with components 
$q_i = \sigma_i$ for $1 \leq i \leq N$, $q_{N+1} =1$, and $q_{N+2} = -1$. 

The determinant in Eq.~(\ref{eq:f_el}) can now be simplified as
\begin{equation} 
\det(1 + \hat{G}_k\hat{U}_k) = 1+ \lambda(k) \langle q | \hat{G}_k | q \rangle 
	\label{eq:Sylvester}
\end{equation}
by Sylvester's identity~\cite{Borue88}.
For the analysis below, we define a reduced temperature 
$T^* \equiv {a}/{l_B} = {4\pi\epsilon_0\epsilon k_{\rm B} Ta}/{e^2}$.

%\section{Results and Discussion}

\begin{figure}
\includegraphics[width=\columnwidth]{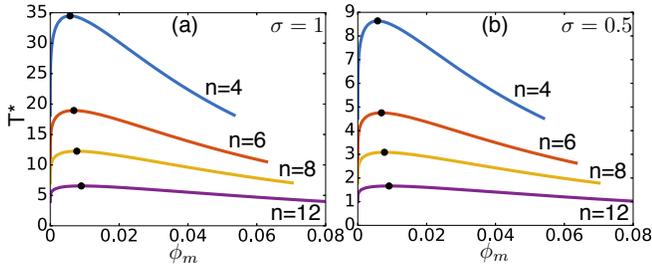} 
\caption{ Phase diagrams of $N=240$
salt-free polyeampholytes with 4, 6, 8, and 12 alternating 
charge blocks of $\sigma_\alpha^\text{block} = (-1)^{\alpha-1}\sigma$;
$\phi_m$ and $T^*$ are, respectively, dimensionless polymer volume ratio 
and temperature (see text).
A dilute and a condensed phase exist, respectively, above
and below each phase-boundary curve. 
(a) $\sigma=1$ and (b) $\sigma=0.5$.
The black dots are the critical points $(\phicr,\Tscr)$. 
Polyampholytes with fewer blocks and stronger $\sigma$
phase separate at higher $T^*$.}
	\label{fig:ps_block_diff}
\end{figure}

We first consider a simple salt-free 
($\rho_s = 0$) solution of polyampholytes consisting of $n$ alternating
charge blocks (labeled by $\alpha$, $\beta=1$,2, $\dots, n$), each with
one charge per $1/\sigma$ monomer, i.e. 
$ \sigma_\alpha^\text{block}= (-1)^{\alpha-1}\sigma$
%per-monomer charge $ \sigma_\alpha^\text{block}= (-1)^{\alpha-1}$ 
and length $L \!=\! N/n$~\cite{Wittmer93}. 
The correlation matrix $\GM(k)$ in this case 
is an $n\times n$ matrix for the blocks, with
%\begin{equation}
%\begin{aligned}
%\GM^\text{block}(k)_{\alpha\alpha} & \frac{1+r}{1-r}L\sigma - \frac{2r(1-r^{L\sigma})}{(1-r)^2}, \\
%=  L \!+\! \frac{72}{(ka)^4 } \left( \frac{1}{6} (ka)^2 L \!-\! 1 + e^{ -\frac{1}{6}L(ka)^2} \right), \\
%\GM^\text{block}(k)_{\alpha\beta} & \frac{r^{(|\alpha-\beta|-1)L\sigma+1}}{(1-r)^2}(1-r^{L\sigma})^2,
%= \frac{36}{(ka)^4} e^{-\frac{1}{6}(|i-j|-1)L(ka)^2} \left( 1-e^{-\frac{1}{6}L(ka)^2}\right)^2, 
%\end{aligned}
%	\label{eq:Gmk_elements}
%\end{equation}
\begin{equation}
\!\!\GM^\text{block}(k)_{\alpha\beta} = \left\{ 
\begin{aligned}
& \frac{1+\zeta}{1-\zeta}L\sigma - \frac{2\zeta(1-\zeta^{L\sigma})}{(1-\zeta)^2} \, , \quad \alpha \!=\! \beta , \\
& \frac{\zeta^{(|\alpha-\beta|-1)L\sigma+1}}{(1-\zeta)^2}(1-\zeta^{L\sigma})^2 \, , \; \alpha \!\neq\! \beta , 
\end{aligned}
\right.
	\label{eq:Gmk_elements}
\end{equation}
where $\zeta = \exp[-(ka)^2/(6\sigma)]$.
%where the $L$ term in the first expression corresponds to the contribution
%from the first ``1'' on the right hand side of Eq.(32) in \cite{Ermoshkin03b},
%and the second expression is for $\alpha\ne\beta$. In the same vein,
%the first $N$ components of $| q \rangle$ is grouped into 
%$n$ components with $q_\alpha  = (-1)^{\alpha-1}$. 
Results for block polyampholytes that are neutral 
($n$ even, $\rho_c =0$; Fig.~\ref{fig:ps_block_diff}) indicate that,
when $N$ is fixed,
the tendency to phase separate decreases (i.e., requires a lower $T^*$)
with increasing $n$ (Fig.~\ref{fig:ps_block_diff}(a)).
This predicted behavior offers insights into Ddx4 behavior (see below)
and is consistent with theoretical findings 
from a charged hard-sphere chain model~\cite{Jiang06} and grand canonical 
Monte Carlo simulations~\cite{Cheong05}.
While a stronger $\sigma$ leads generally to higher
phase separation $T^*$ (and thus higher critical temperature $\Tscr$), 
the critical concentration $\phicr$ is determined mainly
by the block number $n$, and barely by $\sigma$
[cf. curves for the same $n$ in Fig.~\ref{fig:ps_block_diff}(a) and (b)].
When an increasing $n$ arrives at a strictly
alternating polyampholyte with $\sigma = 1/L$, the
contribution of $\GM^\text{block}(k)$ in
$\langle q | \hat{G}_k | q \rangle$, 
\begin{equation}
\frac{1}{N}\langle \sigma | \GM^{\rm block}(k) | \sigma \rangle
= \frac{1-\zeta}{1+\zeta}\sigma +
  \frac{1}{N}\frac{2\zeta[1-(-1)^n \zeta^{N\sigma}]}{(1+\zeta)^2},
\label{eq:sGs_alternating}
\end{equation}
contains the second $O(1/N)$ term that diminishes as $N\to\infty$.
In that limit, the first term leads to $f_{\rm el} \propto \phi_m^2$ after
integration, defining an effective FH parameter
$\chi \propto \sigma^{3/2}/T^{*2}$ for electrostatics~\cite{Wittmer93}. 
When $N$ is finite, the second term enhances phase separation, 
resulting in a $\phicr$ much smaller than the $\phicr =1/\sqrt{N}$ 
predicted by FH theory. For example, for $N=240$, $\sigma = 1$ and 0.5, 
$\phicr$ $=$ 0.0213 and 0.0192, respectively. Both $\phicr$ values 
are much smaller than the FH value of $1/\sqrt{240}=0.06$. 

\begin{figure}
  \includegraphics[width=\columnwidth]{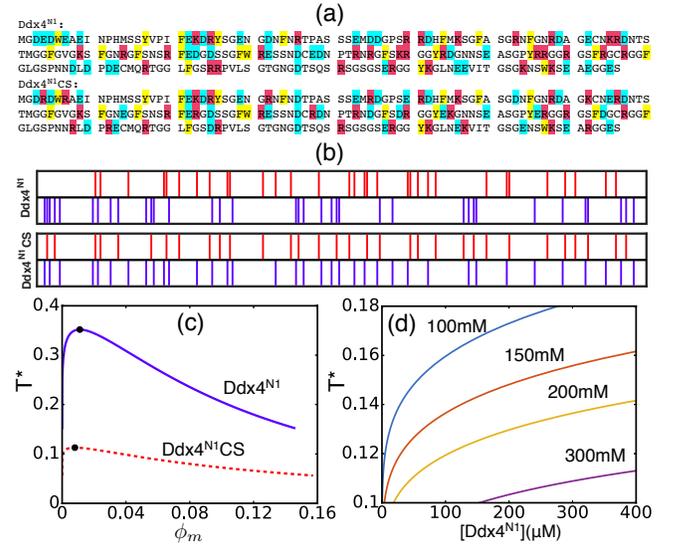} 
  \caption{ (a) The Ddx4$^\text{N1}$ (top) and Ddx4$^\text{N1}$CS (bottom)
 amino acid sequences. Positively (+), negatively ($-$) charged 
(basic and acidic), and aromatic residues are highlighted, respectively, 
in red, turquoise, and yellow. 
(b) Charge patterns of the sequences ($+$: red; $-$: blue) indicate
that the charges exhibit more block-like properties (repeated red 
or repeated blue) 
and are less evenly dispersed in Ddx4$^\text{N1}$ 
than in Ddx4$^\text{N1}$CS (cf. Fig.6A of \cite{Nott15}).
 (c) Theoretical phase diagrams computed by Eq.~(\ref{eq:f_el}) under
salt-free conditions.
(d) Ddx4$^\text{N1}$ phase diagrams [Eq.~(\ref{eq:f_el})] 
for different monovalent salt (NaCl) concentrations. }
	\label{fig:Ddx4_N1-N1CS_compare}
\end{figure}

We now apply the full theory to Ddx4 by considering the exact 
Ddx4$^\text{N1}$ and charged scrambled Ddx4$^\text{N1}$CS sequences
(Fig.~\ref{fig:Ddx4_N1-N1CS_compare}(a)). There are 32 
positively and 36 negatively charged residues in either
sequence (Fig.~\ref{fig:Ddx4_N1-N1CS_compare}(b)), 
the two polyampholytes are thus nearly neutral~\cite{Nott15}. 
We assign $\sigma_i = 1$ to each of the 28 R and 4 K residues, 
$\sigma_i = -1$ to each of the 18 D and 18 E residues,
$\sigma_i = 0$ to all other residues. We then substitute these 
$\{ \sigma_i \}$ values into Eq.~(\ref{eq:f_el}) for $| q \rangle$
(now a $(236+2)$-component vector) to
compute the two sequences' $f_{\rm el}$ and the corresponding
phase diagrams. 
The $\sigma_i$'s are taken to be constant because we are
interested primarily in physiological and/or experimental pH in
the range of 7.0--8.0~\cite{Nott15}, although amino acid charges can 
change significantly if pH variation is larger.
Comparing the two phase boundaries 
in the absence of salt ($\rho_s=0$ but $\rho_c\ne 0$)
indicates that charge scrambling leads to an approximately three fold
decrease in critical temperature (Fig.~\ref{fig:Ddx4_N1-N1CS_compare}(c)),
underscoring once again the importance of charge
pattern in polyampholyte phase behavior and is in qualitative 
agreement with the experimental observation that Ddx4$^\text{N1}$ 
phase separates whereas Ddx4$^\text{N1}$CS does not~\cite{Nott15}.
Interestingly, the predicted $\phicr$ of Ddx4$^\text{N1}$CS is smaller than 
that of Ddx4$^\text{N1}$, which, according to Fig.~\ref{fig:ps_block_diff}, 
suggests that Ddx4$^\text{N1}$CS is akin to a sequence with a weaker $\sigma$
but has longer charge blocks than Ddx4$^\text{N1}$.
Apparently, the sequential proximity of opposite charges 
in Ddx4$^{\rm N1}$CS results in a much weaker effective $\sigma$.

We next explore salt effects on Ddx4$^\text{N1}$ phase behavior by
equating $a^3$ with the volume of a 
single water molecule, in which case $\phi_m$ $=236$[Ddx4$^{\rm N1}$]/55.5 M 
because the molarity of water is 55.5. We consider $\phi_s=$ 
0.0018, 0.0027, 0.0036, and 0.0054, which are calculated in the same 
manner, respectively, for [NaCl]=100, 150, 200, and 300 mM~\cite{Nott15}.
To compare with Fig.~4 of \cite{Nott15}, 
we focus first on the range of [Ddx4$^{\rm N1}$]=0--400$\mu$M. 
Fig.~\ref{fig:Ddx4_N1-N1CS_compare}(d) 
shows a clear decreasing propensity 
(i.e., requiring a lower $T^*$) to phase separate with increasing salt.
This theoretical salt dependence is expected of Ddx4$^\text{N1}$ 
as a nearly neutral 
polyampholyte (whereas phase separation of highly charged polyelectrolytes 
can be promoted by salt~\cite{Dobrynin05}), and is qualitatively 
consistent with experiment~\cite{Nott15}. 

While Fig.~\ref{fig:Ddx4_N1-N1CS_compare}(d) provides a 
conceptual rationalization, it does not match quantitatively
with experimental salt dependence: For [Ddx4$^{\rm N1}$] $\approx 150\mu$M, 
the ratio of absolute 
temperatures at the experimental phase boundaries for [NaCl] $=100$mM  
and $300$mM is about $1.2$ (between $\approx 0^\circ$ 
and $60^\circ$C)~\cite{Nott15}, but the corresponding theoretical
ratio is much higher at $\approx 1.7$. 
This mismatch means that certain ``background'' Ddx4 interactions that are 
less dependent on salt have not been
taken into account by our theory. Indeed, aromatic interactions 
are expected to contribute significantly to Ddx4 phase properties, 
as to other IDP behaviors~\cite{Chen15, Song13}, because a Ddx4$^\text{N1}$ 
mutant with wildtype charge pattern but with 9 of its 14 phenylalanines (F) 
replaced by alanines does not phase separate~\cite{Nott15}. Moreover,
repetitive phenylalanine-glycine (FG) patterns in IDPs are known to 
drive phase separation~\cite{Schmidt15, Brangwynne15}. These observations 
suggest strongly that cation-$\pi$ and $\pi$-$\pi$ stacking interactions 
can play central roles in IDP phase 
separation~\cite{Nott15, Schmidt15, Brangwynne15, Vernon16}. 

\begin{figure}
\includegraphics[width=\columnwidth]{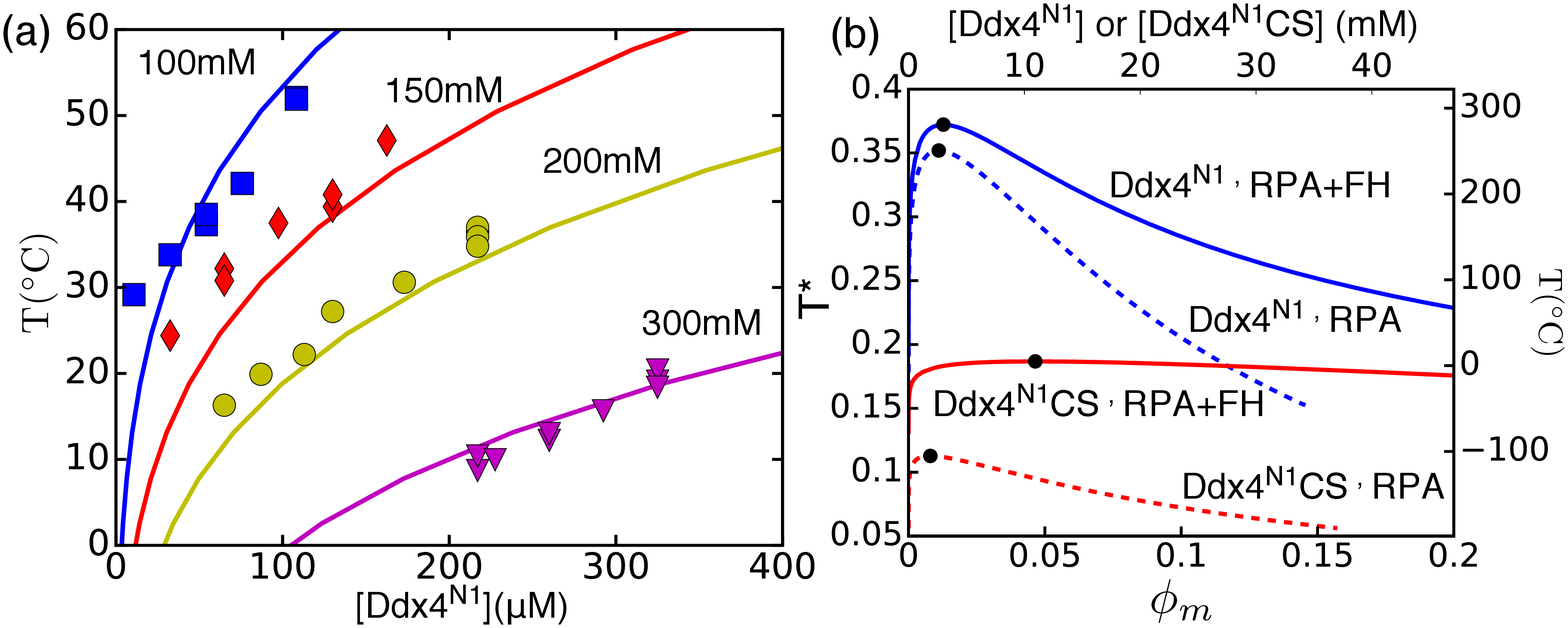}  
\caption{(a) Theoretical phase diagrams of Ddx4$^\text{N1}$ under different 
[NaCl] 
values, computed by RPA theory augmented by FH short-range interactions 
(curves), with fitted $\epsilon=29.5$, $\varepsilon_h = 0.15$, and 
$\varepsilon_s = -0.3$ (see text). Experimental data~\cite{Nott15} 
are included for comparison (symbols; same color code). (b) The salt-free 
phase diagrams of Ddx4$^{\rm N1}$ and Ddx4$^{\rm N1}$CS with and without the 
FH interaction in Eq.(\ref{FH-aug}).} 
	\label{fig:RPAonly_vs_RPA_FH}
\end{figure}

With this consideration, we augment our RPA theory for 
Ddx4$^\text{N1}$ with a mean-field account of $\pi$-interactions, 
which are known to be spatially short-ranged~\cite{Ma97, Meyer03} and
may therefore be formulated, as a first approximation, by a
salt-independent FH term, 
\begin{equation}
f_{\rm FH} = \chi\phi_m(1-\phi_m) = \left(\frac{\varepsilon_h}{T^*}+ 
\varepsilon_s\right)\phi_m(1-\phi_m),
\label{FH-aug}
\end{equation}
where $\varepsilon_h$ and $\varepsilon_s$ are the enthalpic and entropic 
contributions, respectively, to the FH interaction parameter $\chi$. 
To compare theory with experiment quantitatively, values for the monomer 
length scale $a$ and the dielectric constant $\epsilon$ are needed 
to convert $T^*$ to actual temperature.
We take $a$ to be the C$\alpha$-C$\alpha$ distance 3.8\AA, and
let $\epsilon$ be a fitting parameter, as $\epsilon$ 
of an aqueous protein solution can vary widely between 
$\approx$ 2 and 80~\cite{Pitera01, Warshel06}. By treating
$\varepsilon_h$ and $\varepsilon_s$ also as global fitting parameters,
a reasonably good fit is achieved with experimental results
for all four available [NaCl] values~\cite{Nott15} 
(Fig.~\ref{fig:RPAonly_vs_RPA_FH}(a))
by using a value of $\epsilon=29.5$ that is physically plausible
for a mixture of protein ($\epsilon\approx 2$--$4$) and water 
($\epsilon\approx 80$). 

%%%%%%%%%%%% new paragraph %%%%%%%%%%%%%
%%%%%%%%%%%%%%%%%%%%%%%%%%%%%%%%%
%%%%%%%%%%%%%%%%%%%%%%%%%%%%%%%%%
As a self-consistency check, 
we compare the salt-free phase diagrams of 
the wildtype and charge-scrambled Ddx4 sequences computed with the augmented 
FH interaction. Ddx4$^{\rm N1}$CS is then predicted to never 
phase separate at temperature appreciably above 0$^\circ$C
(Fig.~\ref{fig:RPAonly_vs_RPA_FH}(b)), consistent with no observation
of Ddx4$^{\rm N1}$CS phase separation in experiment under
physiological conditions~\cite{Nott15}.
Our model's ability to rationalize the diverse phase behaviors
of Ddx4$^{\rm N1}$ and Ddx4$^{\rm N1}$CS simultaneously suggests that it can 
be applied to predict/rationalize the phase behaviors of other permutations 
of Ddx4$^{\rm N1}$ sequences in the future.
Interestingly, the $\phicr$ of Ddx4$^{\rm N1}$CS 
is dramatically shifted from 0.0080 by the FH term to 0.046, which 
is in the order of the FH critical point $1/\sqrt{236} = 0.065$, 
indicating that the electrostatic interaction 
in Ddx4$^{\rm N1}$CS is weak and its phase behavior is determined 
mainly by the augmented FH interactions. In contrast, the critical point 
of the wildtype Ddx4$^{\rm N1}$ is barely shifted by 
the augmented FH term, implying that its phase behavior is dominated
by electrostatics. 

%%%%%%%%%%%%%%%%%%%%%%%%%%%%%%%%%
%%%%%%%%%%%%%%%%%%%%%%%%%%%%%%%%%
%%%%%%%%%%% new paragraph ends %%%%%%%%%%%

The fitted $\varepsilon_h = 0.15$ and $\varepsilon_s = -0.3$ 
in Fig.~\ref{fig:RPAonly_vs_RPA_FH} correspond 
to a favorable enthalpy of $\Delta H = -0.44$ kcal/mol and an unfavorable 
entropy of $\Delta S = -0.60$ cal/mol K$^{-1}$. 
Notwithstanding the limitations of our model such that part of these 
augmented FH parameters may be needed to correct for the inaccuracies of RPA
itself, the fitted quantities are in line with our 
assumption that they should account approximately for favorable 
$\pi$-interactions. The 
cation-$\pi$ attractive enthalpy between the lysine NH$_4^+$ group 
and an aromatic ring is about 20 kcal/mol in gas-phase ab initio
simulation~\cite{Ma97}, and, because of the 
resonant $\pi$-electron on its guanidinium group~\cite{Marsili08}, 
the attraction between an arginine and an aromatic ring can be even 
stronger, though cation-$\pi$ interaction strength can be weakened
in an aqueous environment~\cite{Gallivan99}. 
Considering there are 32 cations and 22 aromatic rings (14 phenylalanines, 
3 tryptophans, and 5 tyrosines) on Ddx4$^\text{N1}$ and Ddx4$^\text{N1}$CS, 
the effective $\Delta H$ for
cation-$\pi$ interactions in the 
FH parameter may be estimated to be at least of order 
$-20(22\times 32/236^2) =-0.25$kcal/mol (note that $\pi$-$\pi$ energies
and beyond-pairwise multi-body interactions
are not included in this estimation), which matches reasonably
well with the fitted value. Moreover, formation of cation-$\pi$ 
and $\pi$-$\pi$ pairs invariably entail sidechain entropy losses, which
are consistent with a fitted negative value for $\Delta S$.

We notice that curvature of the
phase boundary increases with increasing $\varepsilon_h$. In view of
the experimental trend of decreasing phase boundary
curvature with increasing salt, this observation suggests that
the theory-experiment match may be improved
by allowing $\varepsilon_h$ to decrease slightly with [NaCl].
Such a model may be justifiable because, with increasing salt
concentration, aromatic rings are more likely to bind to surrounding 
salt cations rather than engaging with the positively charged lysines 
and arginines. A detailed analysis of this possibility, however,
is beyond the scope of the present work.

%\section{Conclusion}

In summary, we have developed a general analytical theory for
sequence-specific electrostatic effects on polyampholyte phase separation.
Going beyond mean-field theories that do not consider sequence
information~\cite{Nott15}, our theory provides physical 
underpinnings for experimental Ddx4 behaviors in terms of the charge pattern 
along its sequence and probable $\pi$-interactions. Theory and 
experiment both suggest that an alternating charge pattern that maintains 
reasonably high average positive or negative charge densities over a length 
scale encompassing at least several amino acid residues is required for Ddx4 
phase separation~\cite{Nott15}. It would be instructive to investigate how
this principle is manifested in other IDPs. In any event,
it would be useful to apply our theory to other IDP 
polyampholytes and to generalize the present formulation to incorporate 
sequence-specific non-electrostatic interactions.

\begin{acknowledgments}
We thank Patrick Farber, Lewis Kay, Jianhui Song, and Robert Vernon for 
helpful discussions. This work was supported by a Canadian Cancer Society 
Research Institute grant to J.D.F.-K. and H.S.C. as well as a Canadian 
Institutes of Health Research grant to H.S.C. We are grateful to SciNet 
of Compute Canada for providing computational resources.
\end{acknowledgments}

\bibliographystyle{apsrev4-1}
\bibliography{Lin-FormanKay-Chan-Sep17RR_2016}

%\begin{thebibliography}{99}
%\end{thebibliography}

\end{document}